\documentclass[9pt,twocolumn,twoside]{pnas-new}

\templatetype{pnasresearcharticle} 

\title{A comprehensive understanding of planet formation is required for assessing planetary habitability and for the search for life}

\author[a,b,c]{D\'aniel Apai}
\author[d]{Fred Ciesla}
\author[b,c]{Gijs D. Mulders}
\author[b,c]{Ilaria Pascucci} 
\author[e]{Richard Barry}
\author[f]{Klaus Pontoppidan}
\author[g]{Edwin Bergin}
\author[a,c]{Alex Bixel}
\author[h]{Sean Brittain}
\author[e]{Shawn D. Domagal-Goldman}
\author[i]{Yasuhiro Hasegawa}
\author[j]{Hannah Jang-Condell}
\author[b,c]{Renu Malhotra}
\author[g]{Michael R. Meyer}
\author[a]{Andrew Youdin}
\author[k]{Johanna Teske}
\author[i]{Neal Turner}

\affil[a]{Steward Observatory, The University of Arizona, Tucson, AZ, USA}
\affil[b]{Lunar and Planetary Laboratory, The University of Arizona, Tucson, AZ, USA}
\affil[c]{EOS Team, NASA Nexus for Exoplanet System Science}
\affil[b,d]{University of Chicago, Chicago, IL, USA}
\affil[e]{NASA Goddard Space Flight Center, MD, USA}
\affil[f]{Space Telescope Science Institute, MD, USA}
\affil[g]{University of Michigan, Ann Arbor, MI, USA}
\affil[h]{Clemson University, Clemson, SC, USA}
\affil[i]{Jet Propulsion Laboratory, Pasadena, CA, USA}
\affil[j]{University of Wyoming}
\affil[k]{Hubble Postdoctoral Fellow, Department of Terrestrial Magnetism
Carnegie Institution of Washington, Washington, DC, USA}

\leadauthor{Apai} 

\significancestatement{Astronomical observations of individual exoplanets can only provide an incomplete picture of the properties and processes that determine whether a rocky planet is habitable and/or earth-like. Important statistical context can, however, be gained from understanding how planetary systems form and evolve. Efficient, quick-paced, and comprehensive multi-disciplinary research progress is essential to realize the potential of planet formation/evolution studies, in time to guide our next major steps in the exploration of potentially habitable exoplanets.}

\correspondingauthor{\textsuperscript{2}To whom correspondence should be addressed. E-mail: apai \@ arizona.edu}

\begin{abstract}
Dozens of habitable zone, approximately earth-sized exoplanets are known today. An emerging frontier of exoplanet studies is identifying which of these habitable zone, small planets are actually habitable (have all necessary conditions for life) and, of those, which are earth-like.  Many parameters and processes
influence habitability, ranging from the orbit through detailed
composition including volatiles and organics, to the presence of
geological activity and plate tectonics.   While some properties will
soon be directly observable, others cannot be probed by remote sensing
for the foreseeable future.   Thus, statistical understanding of
planetary systems' formation and evolution is a key supplement to the
direct measurements of planets' properties.  Probabilistically
assessing parameters we cannot directly measure is essential to
reliably assessing habitability, to prioritizing habitable-zone
planets for follow-up, and for interpreting possible biosignatures.
\end{abstract}

\dates{This manuscript was compiled on \today}
\doi{\url{http://eos-nexus.org/whitepapers/}}

\begin{document}

\verticaladjustment{-2pt}

\maketitle
\thispagestyle{firststyle}
\ifthenelse{\boolean{shortarticle}}{\ifthenelse{\boolean{singlecolumn}}{\abscontentformatted}{\abscontent}}{}

\pagebreak
\clearpage

\section*{Habitability and Planetary Properties}

It is foreseen that remote sensing surveys for life beyond the solar system will likely be limited to signatures originating from surface or near-surface life, for the lack of efficient ways to probe sub-surface and deep ocean habitats. For a planetary surface to be {\em habitable} it must not only {allow for liquid water to exist} but the planet and planetary system must provide all conditions necessary for life. Therefore, surface habitability requires the availability of chemical ingredients necessary for life, the presence of an atmosphere, and a relatively stable planetary climate. 

\section*{Relevance of Not Directly Observable Parameters}
Multiple key parameters with direct impact on planetary habitability do not lend themselves to remote-sensing measurements (see Table~\ref{HabFactors}). Constraining or determining these parameters can often be best achieved by understanding the formation and evolution of the planet, and its interactions with the host star, and the evolution of its host planetary system. The most obvious such parameters are the detailed bulk composition (Si, Fe, Mg, C, H, O, N) of the planet, including the planetary volatile and organics budgets. (Although bulk density constraints may allow the identification of {\em clearly non-earth-like} planets, the degeneracies inherent to the equations of state for different possible compositions do not allow identifying habitable planets in general). Another example is geological activity, important for planetary habitability through providing a very large buffer for the atmosphere (acting as a powerful stabilizer against atmospheric losses and climate fluctuations) and for its role in generating a magnetic field (which provides some protection against atmospheric solar wind stripping). Geological activity is extremely difficult to assess via remote-sensing for planets that are broadly earth-like (although see~\cite{Apai2017} for some pathways possible in the distant future). However, the presence or absence of earth-like composition and processes can be predicated probabilistically if the formation history and bulk composition of the planet are reasonably well established. Table~\ref{HabFactors} provides an overview of factors influencing planetary habitability and whether they are best characterized via remote sensing or via constraining the formation and evolution of the given planet.

\begin{table}
\centering
\caption{
Planetary properties directly relevant to habitability.
  While some of these can be measured by remote sensing now or in the
  near future, several of the key parameters will be accessible only
  through our understanding of planetary systems' formation and
  evolution.
  \label{HabFactors}}
\begin{tabular}{lcc}
Properties and Processes & Remote& Formation \\
Constrained by & Sensing & Evolution \\
\midrule
Mass & Y & N  \\
Radius & Y & N  \\
Present-day Irradiation & Y & N \\
Rotation Period& Y & N \\
Present-day Atmospheric Composition & Y & N \\
Present-day Atmospheric/Volatile Loss & Y & Y\\
\midrule
Detailed Bulk Composition (Si/Fe/Ni/O/C) & N & Y \\
Organics/Volatile Inventory (H/C/N) & N & Y \\
Orbital Evolution & N & Y\\
Earth-like Geological Activity & N & Y\\
Past Atmospheric / Volatile Loss & N & Y\\
\bottomrule
\end{tabular}
\end{table}

\section*{Examples for the Importance of Formation and Evolution}
We bring two examples for the importance of understanding planet formation and evolution for establishing a planet's habitability and, through this process, for identifying the ideal target set for biosignature searches. 

{\em Proxima Centaturi b:} The indirectly discovered habitable zone planet Proxima Centauri b 
\cite{AngladaEscude2016} is a good example for the kind of information that should be available for future potentially habitable exoplanets and the evaluation process these and future detections will require.

In short, little information is known about the planet itself: only its orbital period, equilibrium temperature, an $m \sin(i)$ measurement, and a loose constraint on orbital eccentricity. Much more is known, however, about the host star Proxima Centauri and about the population of close-in small planets around M dwarf stars. Many follow-up studies assumed a planet with the face value of the $m \sin(i)$ measurement without considering either the uncertainty of the measured value, the fact that it represents a lower limit, or the fact that the nature of the planet (rocky, icy, gaseous) is undetermined. Similarly, the formation and evolution of the planet is not understood. 

A different approach -- and one that may offer a template for future habitable planet interpretation -- was offered by \cite{BixelApai2017}. In this study probability distributions representing observational constraints (both specific to the individual system as well as derived from population statistics of close-in M dwarf exoplanets) were combined. Indeed, because several of the underlying probability distributions are asymmetric (and some are very broad) the nature of the planet is not straightforward to determine. In fact, that study found a broad probability distribution (with 10-15\% likelihood for Proxima Centauri b being a sub-neptune planet) and an expectation value for its mass that is significantly higher than the measured $m sin(i)$ value and with a very asymmetric uncertainty. 

{\em Future studies of habitable exoplanets will most likely have to interpret the nature of individual planets by combining specific information (on the planet itself, the host star, other planets in the system) with prior distributions of planet properties gained from exoplanet population studies (distributions of orbital elements; mass distribution) and with predicted outcomes from planet formation models (volatile content, possible range of atmospheric loss, migration history, etc.)}.  

{\em TRAPPIST-1 planets:} The recently discovered habitable zone, roughly earth-sized TRAPPIST-1 planets \cite{Gillon2017} offer other examples for the challenges posed by the limited information available on such worlds. Up to three of the planets may be in the present-day habitable zone; however, due to telescope time limitations it is likely that only one or two of them can be followed up spectroscopically by the James Webb Space Telescope. But which one should be targeted? The observed properties of the planets (mass, density) provide important, but limited insights \cite{Grimm2018}. However, considering the properties of the exoplanet population and possible formation/evolution histories of the system is very likely to unveil major differences between the otherwise similar planets (e.g., \cite{Unterborn2017}) and help identify one as the better target to invest JWST time in. Thus, TRAPPIST-1 is another system where the information coming directly from the planets must be complemented by the much greater but more general body of information (context) emerging from planet formation and exoplanet population studies. 

{\bf Allocating major resource, such as telescope time, for a planet to follow up {\em purely} on the basis of directly observed properties (e.g., deepest transit depth or signal-to-noise ratio) is neither a conservative or efficient approach:} the implicit assumption behind such a decision would be that all planets of similar sizes (regardless of differences in their other parameters) are essentially the same --  an assumption we already know to be wrong, as there are clear correlations between planet and system properties (e.g. \cite{Mulders2015a,Mulders2016,Pascucci2018}, EXOPAG SAG13 Report).

\section*{Key Challenges in Planetary System Formation}
In this section we briefly review the key challenges in planet formation and planetary system evolution as they relate to planetary habitability assessments. The list below is not an exhaustive but rather a representative list.

\subsection{Planetesimal formation}
The growth of initial submicron-sized grains to $10^3$ km-sized planetesimals represents a critical, but very poorly understood phase in planet formation. From radioactive dating of iron meteorites (surviving fragments of cores of differentiated minor bodies) it is clear that this growth phase was rapid ($\sim10^5$ yr) in the Solar System \cite{ApaiLauretta2010}, but planetesimals in other planet-forming disks remain undetectable. At least two important challenges have been identified for planetesimal formation: (1) Bodies approaching sizes of 1~m experience strong headwind (gas drag) due to the difference between the Keplerian velocity of the bodies and the sub-keplerian velocity of the partly pressure-supported gas, resulting in rapid in-spiraling and loss of meter-sized objects. (2) For 0.1~m-sized objects collisions tend to be destructive rather than constructive, greatly limiting growth rates (e.g., \cite{BlumWurm2008}). These challenges strongly suggest that planetesimals do not grow via pairwise, constructive collisions, but via another, faster and more efficient process. The internal structure of the primitive Solar System materials (sharply peaked size distribution, lack of units with sizes greater than $10^{-2}$m, evidence for rapid assembly) lends further support to this conclusion. 

Multiple mechanisms have been put forward to explain rapid planetesimal formation, including streaming instability \cite{Johansen2009,BaiStone2010,Carrera2015}, pressure-induced dust traps, eddies, and vortices \cite{Lyra2009,Chambers2010}, and gravitational instability \cite{Youdin2011}.

Understanding planetesimal formation is important for planetary habitability because all solids in rocky planets must pass through the planetesimal stage before being accreted (either early or late). Therefore, the physical process responsible for planetesimal formation will likely also affect the entirety of the solids that will eventually build rocky planets, probably influencing the intrinsic volatile and organics budgets of rocky planets.

\subsection{Protoplanetary disk evolution}

Protoplanetary disks are dynamic objects, through which mass is transported inward and accreted by their stars as part of their final, pre-main sequence evolution.  Evidence for this dynamic evolution is found in astronomical observations, where the infall of material from the disk to the star is observed \citep[e.g.][]{hartmann16} as well as in primitive bodies in our Solar System, such as chondritic meteorites, where materials from very disparate disk environments are mixed together on fine (sub-millimeter) scales \citep[e.g.][]{krot09}.  Together, these lines of evidence suggest that this dynamic evolution occurred over timescales of millions of years, and was fundamental in controlling how the earliest stages of planet formation  proceed.

How the physical properties of a disk change as a result of this dynamic evolution determines the properties of the planets that will eventually emerge.  Whether the mass transport is driven by disk winds \citep{Bai2015} or viscous evolution \citep{hartmann18}, the loss of mass over time, combined with dust growth and settling, will lead to continuously evolving pressures, temperatures, and radiation fluxes within the disk. Further, the transport of mass and redistribution of angular momentum that must accompany it, along with interactions between the gas and dust within the disk, will drive large-scale redistribution of solids prior to their incorporation into planets.  As a result, solids will be exposed to a wide-array of disk environments, with their chemical evolution being determined by the integrated path, and set of environments, that they are exposed to within the disk \citep{cieslasandford12}.  This coupled physical and chemical evolution will ultimately determine what compounds are available as solids to be delivered to planets.

\subsection{Proto-solar nebula in the context of protoplanetary disks}
The solar system planets and minor bodies are a relic of the protoplanetary disks around the young sun, historically referred to as the \textit{solar nebula}. The mass, composition, and location of the planets can be used to reconstruct a \textit{Minimum Mass Solar Nebula} (MMSN) the amount  of material that must at least have been present in the sun’s protoplanetary disk at different heliocentric distances \cite{Hayashi1981}. 
The MMSN provides a reference point for comparing the solar system with protoplanetary disk observations.

The mass and radial distribution of material in protoplanetary disks can be estimated from millimeter-wave observations. 
Spatially resolved observations of millimeter-bright protoplanetary disks indicate that the disk mass in the outer ($\gtrsim 10$ au) regions is consistent with the MMSN \cite{Andrews2010}. The radial distribution of material is typically less centrally peaked than the MMSN (e.g., \cite{Zhang2017}).
Larger surveys at lower spatial resolution indicate that the typical protoplanetary disk around a solar-mass star is less massive than the MMSN with $\sim 10 M_\oplus$ of dust (e.g., \cite{Pascucci2016}).
Observations with ALMA are expected to provide direct constraints on the dust mass and indirect constraints on the gas mass in the giant planet-forming regions ($\sim1-10$ au).

The surface density of the inner disk can also be estimated from exoplanet populations. It is estimated that the \textit{Minimum Mass Extrasolar Nebula} is typically 5 times more massive than the MMSN \cite{ChiangLaughlin2013}. 
Exoplanet populations typically contain more mass than protoplanetary disks at an age of a few million years, indicating that planet formation starts early \cite{NajitaKenyon2014}. An understanding of disk evolution, or direct probes of the early phase of protoplanetary disks, are needed to place the solar system in the context of planet forming regions around other stars.

\subsection{Protoplanetary disk dispersal}
It is well established that the lifetime of protoplanetary disks is a few Myr (e.g., 
\cite{Fedele2010,Ribas2014}) and that by $\sim$10\,Myr most disks do not have enough gas to form Jupiter-mass planets (e.g., 
\cite{Pascucci2006}). Furthermore, with only $\sim$10\% of young disks showing evidence of partial clearing (e.g., \cite{Espaillat2014}) the transition between disk-bearing and disk-less appears to be much shorter than the disk lifetime, only a few $100,000$ years. This dual timescale is currently explained by the combination of two main physical mechanisms: viscous accretion, which dominates the early evolution, and photoevaporation driven by high-energy stellar photons, which takes over accretion when the mass accretion rate drops below the thermal wind mass loss rate (e.g., \cite{Alexander2014} for a review). However, recent non-ideal MHD simulations show very inefficient accretion in the classical MRI-driven viscous scenario while removal of angular momentum by MHD disk winds produces accretion at the observed levels (e.g., \cite{Gressel2015}). This has led to the proposal that that magneto-thermal disk winds alone drive disk evolution and dispersal (e.g., \cite{Bai2015}). Observational diagnostics of thermal and MHD disk winds are growing but cannot yet pin down their relative role in dispersing protoplanetary material (see \cite{ErcolanoPascucci} for a review). 

Disk evolution and dispersal directly impact the formation and evolution of planetary systems.
Disk winds, in combination with dust growth and settling, increase the dust-to-gas mass ratio in the disk midplane, which promotes the formation of planetesimals (e.g. \cite{Carrera2017}). In addition, the preferential removal of H/He rich gas by photoevaporation could result in the gradual enrichment of refractory elements and may be necessary to explain the formation of Jupiter and Saturn with all their constraints (e.g. \cite{AliDib2017}). Gas removal ends giant planet formation and stops planet migration. Depending on the timing of giant planet formation and the magnitude of mass loss rates, star-driven photoevaporation may leave a detectable signature on the observed semi-major axis distribution of giant planets (e.g. \cite{AlexanderPascucci2012},\cite{ErcolanoRosotti2015}). MHD and thermal winds may be also needed to explain the two populations of hot and cool Jupiters \cite{ColemanNelson2016} and influence the migration of planetary embryos, hence impact what type of planets can form in a disk (e.g. \cite{Hasegawa2016}).

\subsection{Volatile and organics delivery to habitable zone planets} 
While 70\% of Earth's surface is covered by water, this critical compound makes up just $\sim$0.1\% of the total mass of the planet.  The low mass suggests that water was delivered by the accretion of more volatile-rich bodies that formed further out in the Solar System, beyond the snow line, where water was able to condense as a solid and be incorporated into planetesimals.  Further, life on Earth requires sufficient delivery of biocritical elements C and N as they are important in biological reactions and atmospheric gases which regulate the temperature and pressure at the surface of the Earth.  Like water, the carriers for the primary carriers for elements are largely expected to have been to volatile to exist as solids where the Earth formed, suggesting delivery of material from more distant regions of the Solar System.

While comets are the most volatile-rich bodies in the Solar System, D/H ratios of water on Earth indicate an asteroidal source of water \citep{alexander12}.  Incorporation of objects from beyond the snow line by planets in the HZ of solar mas star seems to be a natural consequence of planetary accretion, particularly with the aid of giant planets to excite the orbits of bodies in this region \citep{obrien18}.  The efficiency of this delivery appears to decrease as we look at the more common, low-mass stars, as the snow line appears to be located further from the respective habitable zones  and the lower occurrence of giant planets to provide dynamical stirring needed to transport planetesimals across large radial distances \citep{raymond07,ciesla15}.  Planetary processes will also be important in determining the volatile inventory of planets that form, as internal heating from radioactive isotopes, impacts, and subsequent accretion events will drive off volatiles from a planet, with evidence suggesting this occurs throughout the planet formation process \citep{Bergin2015,ciesla15}.

\subsection{Migration of solids, planetary building blocks, and planets.}

The Kepler prime mission has revealed that the occurrence rate of planets in the inner planetary systems (d$<$50~d) is very high, demonstrating that most planetary systems have orders of magnitude more mass in their interiors than the solar system. Furthermore, stellar-mass dependent analysis of the Kepler exoplanet population demonstrated that low-mass stars have more small planets {\em and more} mass in solids on short-period orbits than more massive stars, a trend that runs opposite to 
the stellar-mass dependence of disk masses \cite{Mulders2015b}. These findings strongly argue for the re-distribution of solids in the forming planetary systems: either in the form of the transport of planetary building blocks or via migration of planets. 

\section*{Planet Formation and Exoplanet Characterization}

We will now discuss how constraints (both general and system-specific) from planet formation and evolution will be incorporated into habitable planet characterization.

Given the technical challenge in directly detecting habitable planets, it is anticipated that even the properties of planets that can be derived directly from observations (mass, radius, orbital parameters) will not be well-determined quantities with small and straightforward uncertainties, but will be represented by often-complex probability distributions. 

Assessing the habitability of any specific planet will require assessing a number of factors (see Table~\ref{HabFactors}), each represented by a probability distribution. Two key advantages of this approach is that: i) it provides more realistic treatment of the factors than just working with their expectation values and their uncertainties, and ii) it allows combining constraints specific to the individual planet with probabilistic information derived from exoplanet population studies. For example, in assessing the nature of Proxima Centauri b \cite{BixelApai2017} combined all relevant observed properties (as probability distribution functions) for the planet with priors (planet mass distribution, e.g. \cite{Malhotra2015}) derived from the Kepler sample of close-in planets around M-type host stars. 

This probabilistic approach to describing planetary habitability naturally allows folding even complex probability distributions emerging from planet formation and evolution models. 

\section*{Opportunities and Recommendations}

\subsection*{Opportunities}
The following years and the next decades will bring along major opportunities for progress in planet formation and evolution. In the following we briefly review the most important foreseeable projects.

{\em i) JWST.} NASA's upcoming 6.5m mirror diameter visual/infrared observatory will provide powerful new constraints on planet formation models through a variety of observations, including: a) Dust and gas spectroscopy constraining protoplanetary and debris disk formation and evolution {}; b) Compositional diversity of giant exoplanets (both via planetary transits and eclipses and direct imaging). 

{\em ii) WFIRST.} NASA’s decadal survey recommended, flagship-class Wide-Field Infrared Survey Telescope (WFIRST), now in Phase-A, is scheduled for launch in mid-2025.  WFIRST’s primary exoplanetary mission is a near-infrared wide-field survey to detect and characterize planets from the habitable zone out to unbound planets using gravitational microlensing \cite{Barry2011}. 
This survey will test planet formation models through the anticipated large number statistics of sub-Mars mass planets and planets beyond the snowline. In particular, the WFIRST wide-field instrument will be used to observe 10 fields in the Galactic bulge every 15 minutes for six 72-day seasons during the mission resulting in the microlensing detection of a conservatively estimated 2,600 bound planets, about 20,000 transiting planets and hundreds of unbound planets \cite{Spergel2015}.
WFIRST is uniquely capable of detecting sub-Mars mass planets and is expected to detect about 10--30 super-Earths, with the precise number being of particularly high discriminatory value between existing planet formation model \cite{Muraki2011}.

{\em iii) Next-generation NASA Flagship mission.} Currently four preliminary mission concepts have been selected for pre-study by NASA to aid the evaluation of mission concepts in the 2020 Decadal Survey. Three of these concepts would provide particularly important input for planet formation.

The {\em Large UV Optical Infrared telescope} (LUVOIR) is a $9-15$\,m diameter telescope with 3-4 serviceable instruments  covering from $\sim$200 to 2,500\,nm in imaging and spectroscopy. LUVOIR will have the spatial resolution to probe $\sim$1\,au at the distance of nearby star-forming regions like Taurus. This resolution, combined with improved sensitivity at UV/optical wavelengths, will enable to directly detect accreting protoplanets with masses down to Saturn and image the narrow ($\sim$1-10AU) gaps carved by Neptune mass planets. With 40 times higher sensitivity at UV wavelengths and multi-object spectroscopic capabilities, LUVOIR will efficiently survey the entire Orion complex, trace the evolution and dispersal of the main molecular carriers of C, H, and O during planet assembly, trace molecular and low-ionization metals from disk winds, and determine the absolute abundance patterns in the disk as a function of age. As such it will reveal how the changing disk environment affects the size, location, and composition of planets that form around other stars.

The {\em Origins Space Telescope concept} (OST) is a large (at least 25 m$^2$) 5-600$\mu$m, cold (4K) observatory. With more than 1,000  times higher line sensitivity compared to previous far-infrared observatories, OST is designed to efficiently survey 1,000 planet-forming disks around stars of all masses and evolutionary stage to map their total water content using large numbers of rotational water lines. The same survey will also measure the disk gas masses using the ground-state line of hydrogen deuteride at 112 $\mu$m as a direct proxy for H$_2$. The global volatile content and unbiased molecular gas mass of complete disk populations will be critical inputs to any planet-formation model, which will be very difficult, or impossible, to obtain any other way. OST is therefore directly complementary to ALMA and JWST. The design reference disk survey will cover the 30-600 $\mu$m range, 
opening up a large new discovery space of disk gas tracers beyond water and HD. Finally, OST is anticipated to survey the water D/H ratio in tens of solar system comets, allowing comparisons between volatile content of the solar nebula and that revealed by the disk survey.

\subsection*{Recommendations}

{\em i) Research grants supporting multi-investigator, multi-disciplinary projects.} Due to the multi-disciplinary nature of planet formation and evolution single-investigator grants can only focus on individual facets of the challenge. While these efforts are essential, larger-scale opportunities {\em integrating} knowledge and methodology gained from narrowly focused investigations are necessary to advance the understanding of planet formation to the required levels. 

{\em ii) Research grants supporting focused projects to exploit new datasets emerging from missions not specifically focused on planet formation} (e.g., GAIA, WFIRST, TESS) By combining datasets from multiple sources valuable and novel indirect insights can be gained in planet formation. However, such studies often require a two-step approach, which is not well suited for regular grant opportunities that are very competitive and, therefore, tend to favor projects that are low-risk, focused, and incremental (while still important) over higher-risk, higher-gain projects and projects that are integrative in nature. 

{\em iii) Integrative communication channels.} Integrating knowledge, disseminating results, and coordinating progress remains a challenge for the multi-disciplinary community studying planet formation. For example, cosmochemists studying the volatile and organics inventory of the proto-solar system rarely follow the developments in the exoplanet population statistics and vice versa. Although important progress has been made in better utilizing video communication and social media in everyday interactions and in disseminating results, it still often takes many years for new knowledge to propagate through the community. Modern, multi-disciplinary communications channels should be established to increase the efficiency of inter-disciplinary knowledge transfer.

{\em iv) Accelerated incubation process: Faster-paced ideas-to-publications pipeline.} Currently, the typical idea-to-project timescale is about 4-5 years (1~yr to receive funding, 0.5 yr for recruitment, 2-3 yr research/publication). This means only one complete cycle during JWST's minimum lifetime and only about two complete cycles before the first light of the ELTs. Even more concerning is the timeline for truly innovative or paradigm-changing ideas: by their nature these ideas may not be immediately valued in the peer-review process and they may also be inherently more risky. {\em It is important to ensure that novel ideas can be tested quickly.}
While three-year grants leading to multiple publications should remain the cornerstone of research funding, it is important to explore how the ideas-to-publication timeline could be shortened. An obvious possibility is to offer quick turn-around one-year seed grants that require significantly less overhead to write and to evaluate. Successful projects would then have 9-12 months funding period to demonstrate feasibility or reduce risks before competing for a larger and longer grant. A similar approach is adopted by a variety of NASA funding opportunities already.

\section*{Summary}

The key points of our white paper are summarized as follows:

(i) Several key properties of habitable zone planets that are necessary for planetary habitability are not directly observable.

(ii) Evaluating planetary habitability and interpreting biosignatures will require both a contextual and system-level understanding of planet formation and evolution.

(iii) Key challenges to understanding habitable planets  through planet formation include: planetesimal formation, protoplanetary disk evolution and dispersal, the interpretation of the proto-solar nebula and the solar system in the context of other forming planetary systems, volatile and organics delivery to forming planets, the migration of planetary building blocks and planets, atmospheric loss and  atmospheric replenishment. 

Developing the understanding of planet formation to the level required for selecting targets for habitable planet characterization experiments, interpreting their results, and - in particular - for correctly interpreting biosignatures will require changes in the way planet formation studies are funded and in the ways the community is connected:

(a) Focused single-investigator grants must be complemented by large-scale, multi-investigator grants to integrate the multi-disciplinary  research on planet formation.

(b) Interdisciplinary communication remains a limiting factor in the spreading of ideas and the launch of new projects. More efficient use of modern collaborative tools and social media should further increase information flow between the disciplines.

(c) A shorter ideas-to-papers timeline to ensure more rapid progress and quicker exploration of new ideas would be enabled by the introduction of quick turn-around seed grants and/or two-step grants


\subsection*{References}






\bibliography{pnas-sample}

\begin{thebibliography}{10}

\bibitem{Apai2017}
{Apai} D, et~al. (2017).
\newblock {\em ArXiv e-prints}.

\bibitem{AngladaEscude2016}
{Anglada-Escud{\'e}} G, et~al. (2016).
\newblock {\em Nature} 536:437--440.

\bibitem{BixelApai2017}
{Bixel} A, {Apai} D (2017).
\newblock {\em Astrophysical Journal Letters} 836:L31.

\bibitem{Gillon2017}
{Gillon} M, et~al. (2017).
\newblock {\em Nature} 542:456--460.

\bibitem{Grimm2018}
{Grimm} SL, et~al. (2018).
\newblock {\em ArXiv e-prints}.

\bibitem{Unterborn2017}
{Unterborn} CT, {Desch} SJ, {Hinkel} NR, {Lorenzo}, Jr A (2017).
\newblock {\em ArXiv e-prints}.

\bibitem{Mulders2015a}
{Mulders} GD, {Pascucci} I, {Apai} D (2015).
\newblock {\em Astrophys. J.} 798:112.

\bibitem{Mulders2016}
{Mulders} GD, {Pascucci} I, {Apai} D, {Frasca} A, {Molenda-{\.Z}akowicz} J
  (2016).
\newblock {\em Astron. J.} 152:187.

\bibitem{Pascucci2018}
{Pascucci} I, {Mulders} GD, {Gould} A, {Fernandes} R (2018).
\newblock {\em ArXiv e-prints}.

\bibitem{ApaiLauretta2010}
{Apai} DA, {Lauretta} DS (2010) {\em {Protoplanetary Dust: Astrophysical and
  Cosmochemical Perspectives}}.
\newblock pp. 128--160.

\bibitem{BlumWurm2008}
{Blum} J, {Wurm} G (2008).
\newblock {\em Annual Rev. Astron. Astrophys.} 46:21--56.

\bibitem{Johansen2009}
{Johansen} A, {Youdin} A, {Mac Low} MM (2009).
\newblock {\em Astrophysical Journall} 704:L75--L79.

\bibitem{BaiStone2010}
{Bai} XN, {Stone} JM (2010).
\newblock {\em Astrophysical Journal} 722:1437--1459.

\bibitem{Carrera2015}
{Carrera} D, {Johansen} A, {Davies} MB (2015).
\newblock {\em Astron. \& Astrophys.} 579:A43.

\bibitem{Lyra2009}
{Lyra} W, {Johansen} A, {Klahr} H, {Piskunov} N (2009).
\newblock {\em Astron. \& Astrophys.} 493:1125--1139.

\bibitem{Chambers2010}
{Chambers} JE (2010).
\newblock {\em Icarus} 208:505--517.

\bibitem{Youdin2011}
{Youdin} AN (2011).
\newblock {\em Astrophysical Journal} 731:99.

\bibitem{hartmann16}
{Hartmann} L, {Herczeg} G, {Calvet} N (2016).
\newblock {\em Annual Reviews of Astronomy \& Astrophysics} 54:135--180.

\bibitem{krot09}
{Krot} AN, et~al. (2009).
\newblock {\em Geochimica et Cosmochimica Acta} 73:4963--4997.

\bibitem{Bai2015}
{Bai} XN (2015).
\newblock {\em Astrophys. Journal} 798:84.

\bibitem{hartmann18}
{Hartmann} L, {Bae} J (2018).
\newblock {\em Monthly Notices of the Royal Astronomical Society} 474:88--94.

\bibitem{cieslasandford12}
{Ciesla} FJ, {Sandford} SA (2012).
\newblock {\em Science} 336:452.

\bibitem{Hayashi1981}
{Hayashi} C (1981).
\newblock {\em Progress of Theoretical Physics Supplement} 70:35--53.

\bibitem{Andrews2010}
{Andrews} SM, {Wilner} DJ, {Hughes} AM, {Qi} C, {Dullemond} CP (2010).
\newblock {\em Astrophys. Journal} 723:1241--1254.

\bibitem{Zhang2017}
{Zhang} K, {Bergin} EA, {Blake} GA, {Cleeves} LI, {Schwarz} KR (2017).
\newblock {\em Nature Astronomy} 1:0130.

\bibitem{Pascucci2016}
{Pascucci} I, et~al. (2016).
\newblock {\em Astrophys. Journal} 831:125.

\bibitem{ChiangLaughlin2013}
{Chiang} E, {Laughlin} G (2013).
\newblock {\em Monthly Not. Royal Astr. Soc.} 431:3444--3455.

\bibitem{NajitaKenyon2014}
{Najita} JR, {Kenyon} SJ (2014).
\newblock {\em Monthly Not. Royal Astr. Soc.} 445:3315--3329.

\bibitem{Fedele2010}
{Fedele} D, {van den Ancker} ME, {Henning} T, {Jayawardhana} R, {Oliveira} JM
  (2010).
\newblock {\em Astron. \& Astrophys.} 510:A72.

\bibitem{Ribas2014}
{Ribas} {\'A}, {Mer{\'{\i}}n} B, {Bouy} H, {Maud} LT (2014).
\newblock {\em Astron. \& Astrophys.} 561:A54.

\bibitem{Pascucci2006}
{Pascucci} I, et~al. (2006).
\newblock {\em Astrophysical Journal} 651:1177--1193.

\bibitem{Espaillat2014}
{Espaillat} C, et~al. (2014).
\newblock {\em Protostars and Planets VI} pp. 497--520.

\bibitem{Alexander2014}
{Alexander} R, {Pascucci} I, {Andrews} S, {Armitage} P, {Cieza} L (2014).
\newblock {\em Protostars and Planets VI} pp. 475--496.

\bibitem{Gressel2015}
{Gressel} O, {Turner} NJ, {Nelson} RP, {McNally} CP (2015).
\newblock {\em Astrophys. Journal} 801:84.

\bibitem{ErcolanoPascucci}
{Ercolano} B, {Pascucci} I (2017).
\newblock {\em Royal Society Open Science} 4:170114.

\bibitem{Carrera2017}
{Carrera} D, {Gorti} U, {Johansen} A, {Davies} MB (2017).
\newblock {\em Astrophysical Journal} 839:16.

\bibitem{AliDib2017}
{Ali-Dib} M (2017).
\newblock {\em Monthly Notices of Roy. Ast. Soc.} 464:4282--4298.

\bibitem{AlexanderPascucci2012}
{Alexander} RD, {Pascucci} I (2012).
\newblock {\em Monthly Notices of Roy. Ast. Soc.} 422:L82--L86.

\bibitem{ErcolanoRosotti2015}
{Ercolano} B, {Rosotti} G (2015).
\newblock {\em Monthly Notices of Roy. Ast. Soc.} 450:3008--3014.

\bibitem{ColemanNelson2016}
{Coleman} GAL, {Nelson} RP (2016).
\newblock {\em Monthly Notices of Roy. Ast. Soc.} 460:2779--2795.

\bibitem{Hasegawa2016}
{Hasegawa} Y (2016).
\newblock {\em Astrophysical Journal} 832:83.

\bibitem{alexander12}
{Alexander} CMO, et~al. (2012).
\newblock {\em Science} 337:721.

\bibitem{obrien18}
{O'Brien} DP, {Izidoro} A, {Jacobson} SA, {Raymond} SN, {Rubie} DC (2018).
\newblock {\em ArXiv e-prints}.

\bibitem{raymond07}
{Raymond} SN, {Scalo} J, {Meadows} VS (2007).
\newblock {\em Astrophysical Journal} 669:606--614.

\bibitem{ciesla15}
{Ciesla} FJ, {Mulders} GD, {Pascucci} I, {Apai} D (2015).
\newblock {\em Astrophysical Journal} 804:9.

\bibitem{Bergin2015}
{Bergin} EA, {Blake} GA, {Ciesla} F, {Hirschmann} MM, {Li} J (2015).
\newblock {\em Proceedings of the National Academy of Science} 112:8965--8970.

\bibitem{Mulders2015b}
{Mulders} GD, {Pascucci} I, {Apai} D (2015).
\newblock {\em Astrophys. J.} 814:130.

\bibitem{Malhotra2015}
{Malhotra} R (2015).
\newblock {\em Astrophys. J.} 808:71.

\bibitem{Barry2011}
{Barry} R, et~al. (2011) {The exoplanet microlensing survey by the proposed
  WFIRST Observatory} in {\em Techniques and Instrumentation for Detection of
  Exoplanets V}, Proc. SPIE.
\newblock Vol.{} 8151, p. 81510L.

\bibitem{Spergel2015}
{Spergel} D, et~al. (2015).
\newblock {\em ArXiv e-prints}.

\bibitem{Muraki2011}
{Muraki} Y, et~al. (2011).
\newblock {\em Astrophysical Journal} 741:22.

\end{thebibliography}

\end{document}